\documentclass[aps,prl, twocolumn,amsmath,amssymb,superscriptaddress,floatfix]{revtex4-1}

\usepackage{amsmath}
\usepackage{xcolor}
\usepackage{graphicx}
\usepackage{dcolumn}
\usepackage{bm}
\usepackage{verse}

\newcommand{\CNN}{Universit{\'e} Paris-Saclay, CNRS, Centre de Nanosciences et de Nanotechnologies, 91120 Palaiseau, France}
\newcommand{\IMEC}{IMEC, Kapeldreef 75, B-3001 Leuven, Belgium}

\begin{document}


\title {Quenching stochasticity in spin-torque induced domain wall motion}
\title {Stochastic processes in magnetization reversal involving domain wall motion in magnetic memory elements}

\author{Paul Bouquin}
\affiliation{\CNN}
\affiliation{\IMEC}
\author{Joo-Von Kim}
\affiliation{\CNN}
\author{Olivier Bultynck}
\author{Siddharth Rao}
\author{Sebastien Couet}
\author{Gouri Sankar Kar}
\affiliation{\IMEC}
\author{Thibaut Devolder}
\email{thibaut.devolder@u-psud.fr}
\affiliation{\CNN}

\date{\today}

\begin{abstract}
We show experimentally through time-resolved conductance measurements that magnetization reversal through domain wall motion in sub-100 nm diameter magnetic tunnel junctions is dominated by two distinct stochastic effects. The first involves the incubation time related to domain wall nucleation, while the second results from stochastic motion in the Walker regime. Micromagnetics simulations reveal several contributions to temporal pinning of the wall near the disk center, including Bloch point nucleation and wall precession. We show that a reproducible ballistic motion is recovered when Bloch and Néel wall profiles become degenerate in energy in optimally sized disks, which enables quasi-deterministic motion.

\end{abstract}

\maketitle

The reversal of magnetization in nanostructures is a challenging problem for fundamental studies and technological applications. Beyond the coherent reversal mode in which all moments precess in unison as the magnetization switches from one state (“up”) to another (“down”), which is predicted  \cite{sun_spin-current_2000, butler_switching_2012, pinna_thermally-assisted_2013, tomita_unified_2013} for lateral dimensions below $\approx$25 nm but rarely observed in practice \cite{sun_spin-torque_2013, hahn_time-resolved_2016, devolder_size_2016}, the possibility of intermediate states involving nonuniform magnetic textures makes quantitative prediction of switching thresholds and switching times a difficult task. The issue is exacerbated at finite temperatures, where thermal fluctuations render the reversal process stochastic. This is particularly problematic for applications in information storage, where deterministic switching is sought \cite{khvalkovskiy_basic_2013}. 

For perpendicularly-magnetized thin film disks with lateral dimensions greater than 25 nm, the reversal mode following the minimum energy path is predicted to involve the nucleation and propagation of a magnetic domain wall \cite{you_switching_2014, chaves-oflynn_thermal_2015, sampaio_disruptive_2016, bouquin_size_2018, desplat_entropy-reduced_2020}. Such modes are therefore subject to stochastic effects in both the process of nucleation, which generally occurs at edge boundaries, and in the process of propagation as the wall sweeps across the nanostructure. Since dipolar fields are nonuniform across such finite-sized systems, thermal fluctuations can induce a variety of phenomena that can transform the wall structure during reversal.

In this Letter, we present experimental evidence of strong stochastic contributions to the free layer reversal in circular magnetic tunnel junctions under Spin-Transfer Torques (\cite{slonczewski_currents_2005}, STT). Results from time-resolved measurements are interpreted with the aid of micromagnetics simulations and analytical modeling, which show that large-angle precession of the magnetization within the wall, and the nucleation of Bloch points can contribute to temporal pinning effects observed. We show how such effects can be mitigated when Bloch and Néel wall structures become degenerate in energy; this suppresses most of the variability in the domain wall propagation dynamics.

\begin{figure}[t!]
\begin{center}
\includegraphics[width=7. cm]{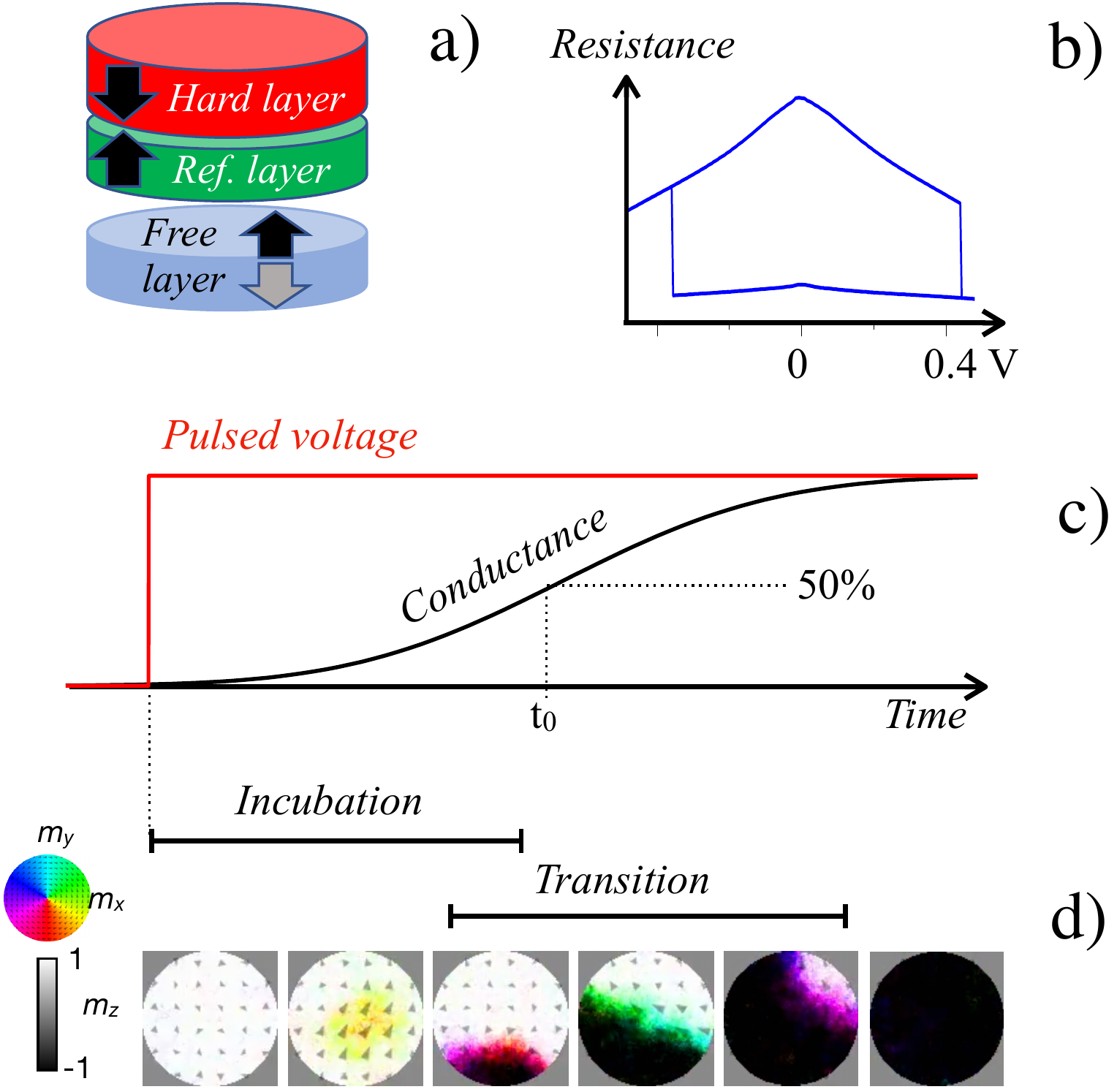}
\caption{a): Device geometry. b) Resistance versus Voltage hysteresis loop. c) Sketch of the experimental procedure. d) Snapshots of some micromagnetic configurations during the simulated reversal of a 80 nm device at 300 K for a spin current of $5.4\times10^{10}~\textrm{A/m}^2$.}
\label{fig1}
\end{center}
\end{figure}
The devices studied are presented in Fig.~1(a): they are CoFeB/MgO-based magnetic tunnel junctions (MTJs) with reference and hard layers organized in a standard synthetic antiferromagnet configuration. The MTJ is switchable by STT at zero magnetic field  [Fig. 1(b)]. It is specifically optimized \cite{devolder_back-hopping_2020, devolder_material_2018} to ensure easy domain wall propagation within the free layer. 
The experimental focus is on a device of diameter 100 nm, typical of the behavior in the interval of investigated sizes (70-100 nm). 

Our set-up applies fast rising voltage steps with maximally flat plateaus and then monitors the time-resolved device conductance [Fig.~1(c)]. 
As a response, the device incubates during a variable delay and then its conductance switches abruptly. We fit the conductance waveforms with the ansatz $\textrm{erf} [(t-t_0) / \tau]$ to define an incubation delay $t_0$ and a transition time $\tau$, which is thus the 24-76\% rise time of the conductance. Micromagnetic simulations indicate that the transition time corresponds to the sweeping of a domain wall (DW) through the device [Fig.~1(d)].

\begin{figure}[t!]
\includegraphics[width=7.5 cm]{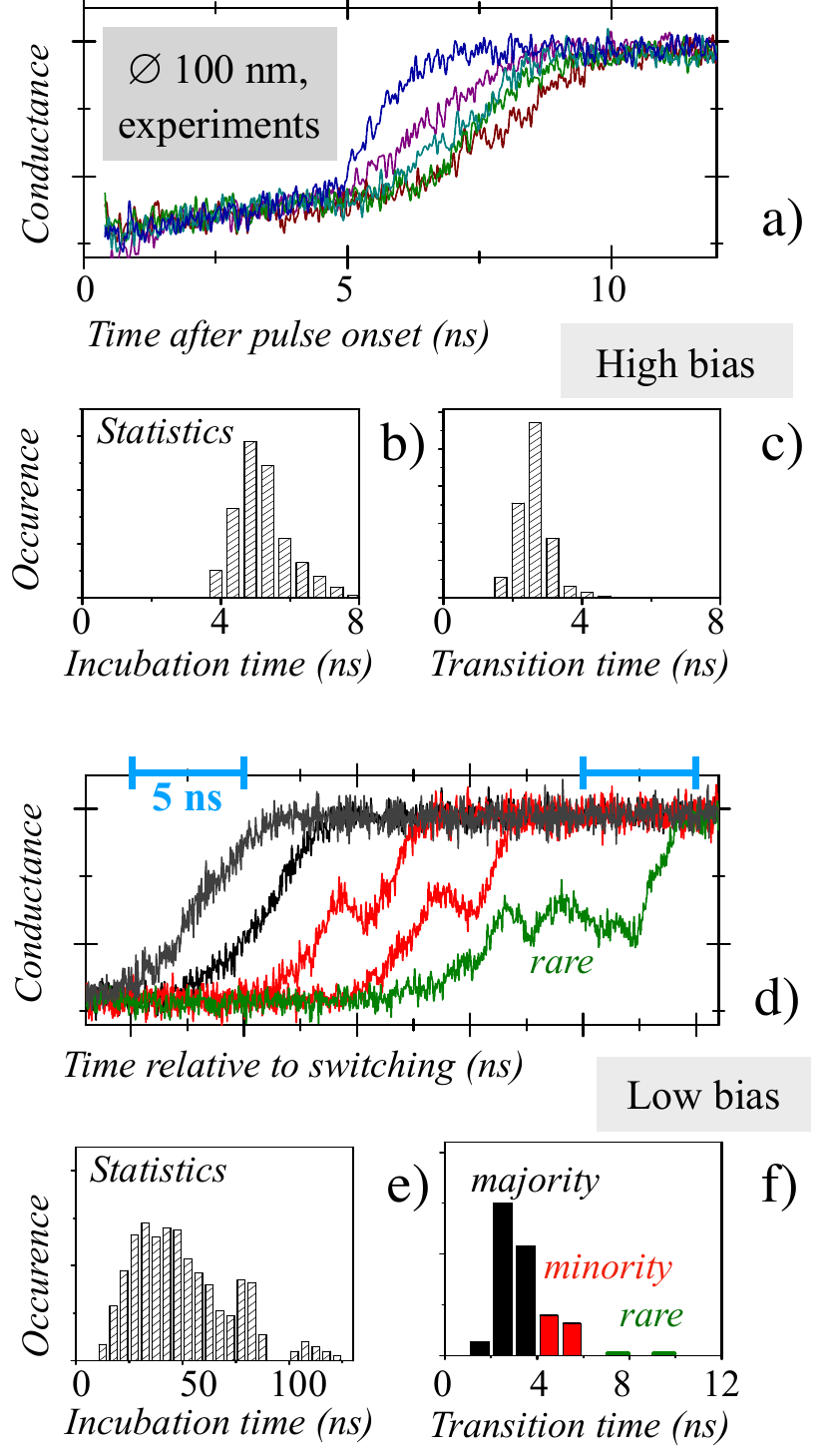}
\caption{Experimental signatures of the switching as a result of voltage pulses of (a-c): 630 mV and (d-f): 500 mV on a 100 nm device. At low bias [panel d)] the curves are horizontally offset to remove incubation delay and reveal the three classes of switching events: ballistic crossing of the midway conductance (black),  crossing with one single pronounced oscillation (red) and multiple crossing (green).}
\label{fig2}

\end{figure}

Results from time-resolved electrical measurements of the switching are shown in Fig. 2.
In the first 5 ns after the pulse onset, the conductance rises asymptotically, likely as a consequence of Joule heating. The subsequent evolution reflects the magnetic moment $\langle m_z \rangle$ of the free layer, which noticeable fluctuations of the incubation delay and of the transition time. Two distinct switching regimes are observed, depending of the magnitude of STT. At high bias [Fig.~\ref{fig2}(a)], the conductance waveforms are monotonic, with ns-scale incubation delay delays and transition times. In the given example, the incubation lasts in average $\langle t_0 \rangle =5.6$ ns and is slightly skewed to higher values [Fig.~\ref{fig2}(b)]. Conversely, the distribution of transition times is rather symmetric about its average value $\langle \tau \rangle=2.5~\textrm{ns}$ [Fig.~\ref{fig2}(c)].

When reducing the STT to just above the quasi-static switching threshold (low bias regime), the dynamics slows down while getting progressively more complex, particularly when the conductance is approximately at midway between the initial and final states [Fig.~\ref{fig2}(d)]. The distribution of transition times becomes notably asymetric. Three categories of switching events can be identified  [Fig.~\ref{fig2}(d)]. In most events (black curves), the conductance still evolves monotonically with transition times typically of 3 to 4 ns. We will see that these "ballistic" curves correspond to a scenario in which a domain wall sweeps rather regularly through the device. For a minority of events (20\% probability, red curves), a pronounced oscillation is observed: the conductance first passes above the midway value, then it recesses for $1.3\pm0.1~\textrm{ns}$ until it finally rises again till saturation. The transition time is longer, typically between 4 and 6 ns.  We shall refer to these events as "central oscillation" events as we will see that they arise when the wall stops on either sides of the disk center during an oscillation about the disk center. 
Finally, in rare occasions ($\approx$2\% probability, green curves) the midway conductance is crossed multiple times and the transition time exceeds 7 ns. These last events shall be referred to as the "multiple-swing" events. We will see that they occur when a Bloch line appears in the wall while it is slow the central region of the disk.  These interpretations are based on the forthcoming micromagnetic simulations that indeed reproduce the key experimental observations.

Our micromagnetics simulations rely on the mumax3 code which solves the Landau-Lifshitz-Gilbert equation with spin-transfer torques using the finite difference method \cite{vansteenkiste_design_2014}. The 80-nm diameter, 2-nm thick circular disk is discretized with 96$\times$96$\times$1 cells, in a material \cite{bouquin_size_2018} of saturation magnetization of $M_s=1.2 \ \mathrm{MA}/\mathrm{m}$, exchange stiffness $A_\mathrm{ex}=20 \ \mathrm{pJ}/ \mathrm{m}$,  perpendicular anisotropy constant $K_u= 1.18~\textrm{MJ/m}^3$, and Gilbert damping $\alpha=0.01$. The simulations are conducted assuming a temperature of 300 K, which is accounted for through the inclusion of white noise in the effective field, and the Langevin dynamics is solved using an adaptive-time step method \cite{leliaert_adaptively_2017}. STT is accounted for by a symmetric Slonczewski term, where we have assumed unit spin polarization for simplicity. 
\begin{figure}[t!]
\begin{center}
\includegraphics[width=7.5 cm]{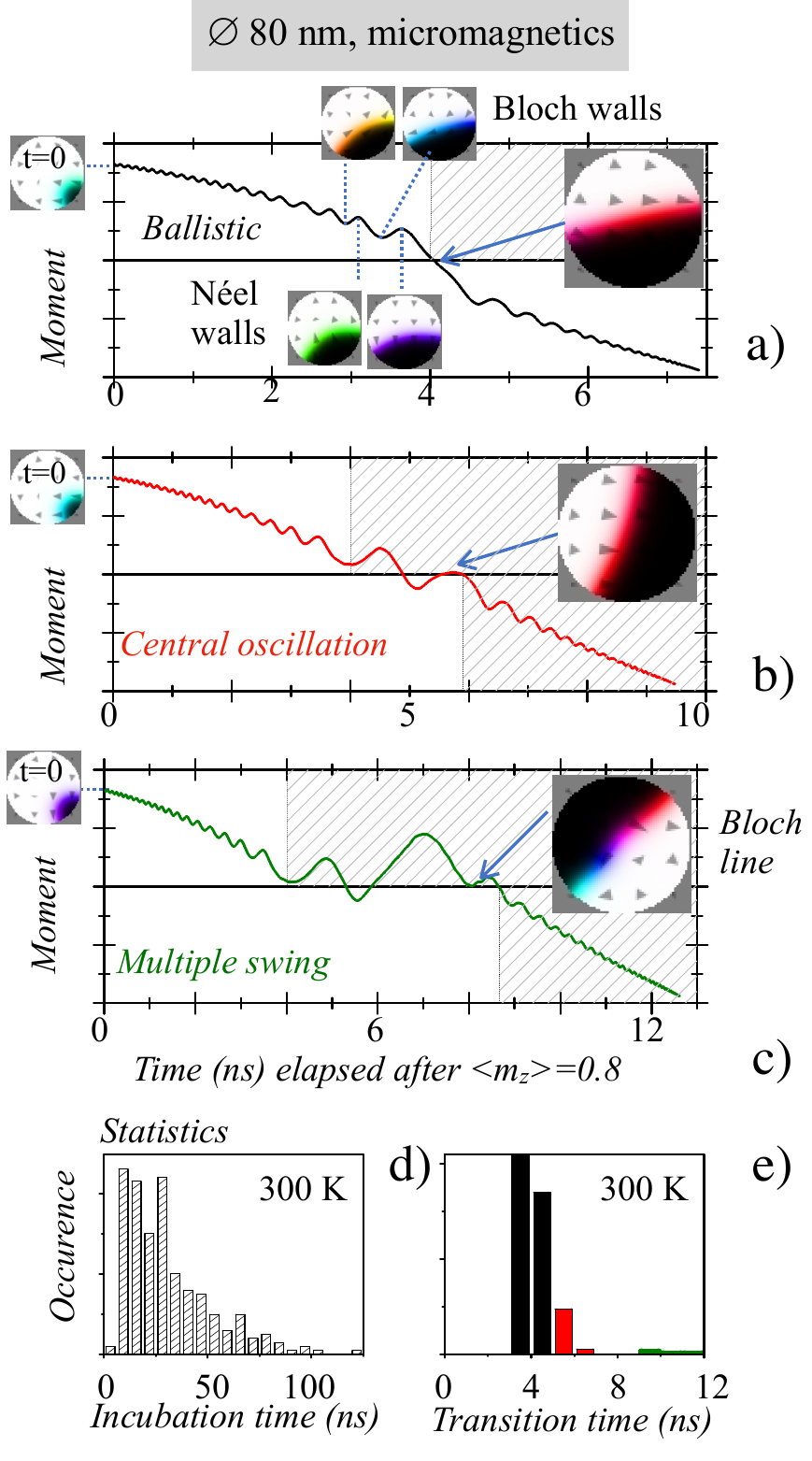}
\caption{Simulated DW dynamics within an 80 nm disk in a low bias situation. (a-c): Examples of dynamics at T=0 K 
for 3 initial states that differ only in wall tilt. The dashed areas underline the switching times. (d) and (e): Statistics of the incubation delays and transition times for a full switching simulated at T=300 K at $3.5\times10^{10}~\textrm{A/m}^2$. } 

\label{fig3}
\end{center}
\end{figure}


Fig.~\ref{fig3} presents the simulated reversal curves for the low bias regime. As expected, the reversal takes place through the nucleation of a 180-degree  wall followed by its propagation. The nucleation is preceded by the growth of a fluctuating “droplet” of precessing moments, which leads to nucleation when it encounters the disk edge. The wall then drifts across the device by a fast back-and-forth oscillatory motion that is reminiscent \cite{devolder_time-resolved_2016} of the wall precessional motion that occurs above the Walker breakdown. The three categories of curves observed experimentally, namely “majority”, “minority”, and “rare”, are reproduced with qualitatively similar probabilities in the simulations.
The "rare" events exhibit outlier transition times, in the 8-10 ns range, with multiple oscillations of the wall near the midway moment; they occur when a Bloch line nucleates within the wall while the latter is near the device center [see inset in Fig.~\ref{fig3}(c)]. The other events have faster transition times and the magnetization within the wall stays rather uniform. The "majority" case  correspond to the ballistic propagation of the wall across the device center, leading to either a linear shape or a faint inflection point when $\langle m_z(t) \rangle =0$. In the "minority" cases, the curves exhibit a pronounced central oscillation when $\langle m_z \rangle \approx0$. This reflects the wall performing two pauses at either sides of the device diameter.

\begin{figure}
\begin{center}
\includegraphics[width=7.5 cm]{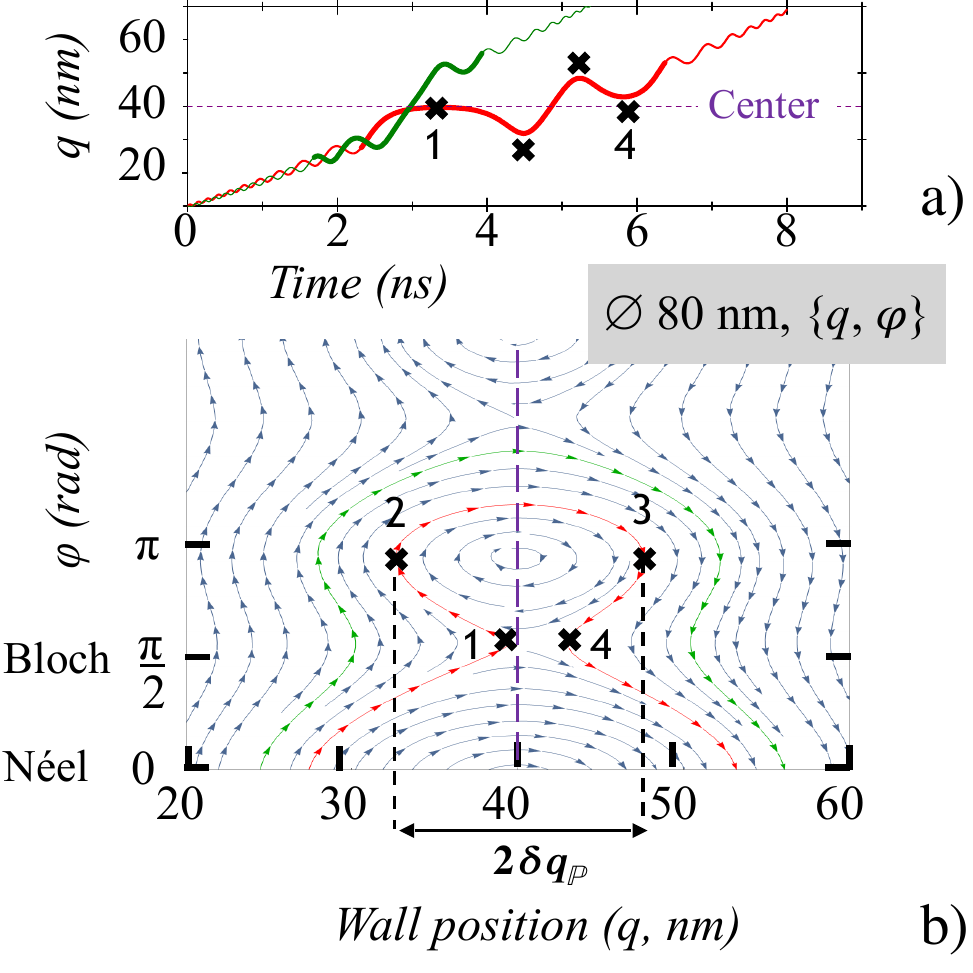}
\caption{Domain wall dynamics within a 80 nm disk in the $\{q,\phi\}$ model at low bias $(4.5\times10^{10}~\textrm{A/m}^2)$. a) Wall trajectories for walls initialized at $q$=10 nm with initial tilts of 48 deg. (green) and 91 deg. (red). b) Corresponding trajectories in the $\{q,\phi\}$ space. The red contour between the labels 1,2,3 and 4 is very close to the frontier of the retention pond $\mathbb P$.} 
\label{fig4}
\end{center}
\end{figure}

Further insight into these processes can be gleaned by an extension to the one-dimensional $\{q, \phi\}$ model~\cite{thiaville_domain_2004} of domain wall dynamics, where the motion is parametrized entirely by the position $q$ of a straight wall and the internal wall angle $\phi$ (the "tilt"), which describes the chirality (Bloch or Néel) of the wall structure. The extension comprises accounting for the spatially non-uniform potential for the wall dynamics, which captures the fact that the disk center appears as an energy barrier to overcome, where the barrier height is related to the additional cost in domain wall energy required to extend it laterally across the disk. The equations of motion become:
\begin{gather}
-\Dot{\phi} + \alpha \frac{\Dot{q}}{\Delta} = - \gamma_0 \left[H_z+H_\mathrm{d}(q)+H_\mathrm{str}(q, \phi)\right]
\label{model3a} \\
\frac{\Dot{q}}{\Delta}+ \alpha \Dot{\phi}  = \gamma_0 \frac{H_{N\leftrightarrow B}}{2} \sin{2\phi} + \sigma j, 
\label{model3b}
\end{gather}
where $\pi \Delta$ is the wall width and $\sigma j$ the magnitude of STT \cite{bouquin_size_2018}. $H_z$ is the applied field. $H_d$ is the stray field of the two domains, $H_{N\leftrightarrow B}$ (typically 30 mT) is the in-plane field that would be needed to transform a Bloch DW into a N\'eel DW \cite{mougin_domain_2007}. To describe the energy barrier in $q$ and $\phi$ directions, we have defined the two contributions to the wall stretch field:
\begin{equation}
H_\mathrm{str}(q) = \left(H_k^\textrm{eff}+ \frac{H_{N\leftrightarrow B}}{2} \cos^2{\phi} \right) \frac{\Delta}{\ell_\mathrm{DW}} \frac{\partial \ell_\mathrm{DW}}{\partial q}
\label{stretchfield}
\end{equation}
where $\ell_\mathrm{DW}(q) = \frac{1}{\Delta} \int ^R _{-R} \mathrm{d}x \ \mathrm{sech}^2(\frac{x-q}{\Delta}) \sqrt{R^2-q^2}$ is the effective length of the straight wall and $R$ is the disk radius. 

\begin{figure}[t!] 
\begin{center}
\includegraphics[width=7.5 cm]{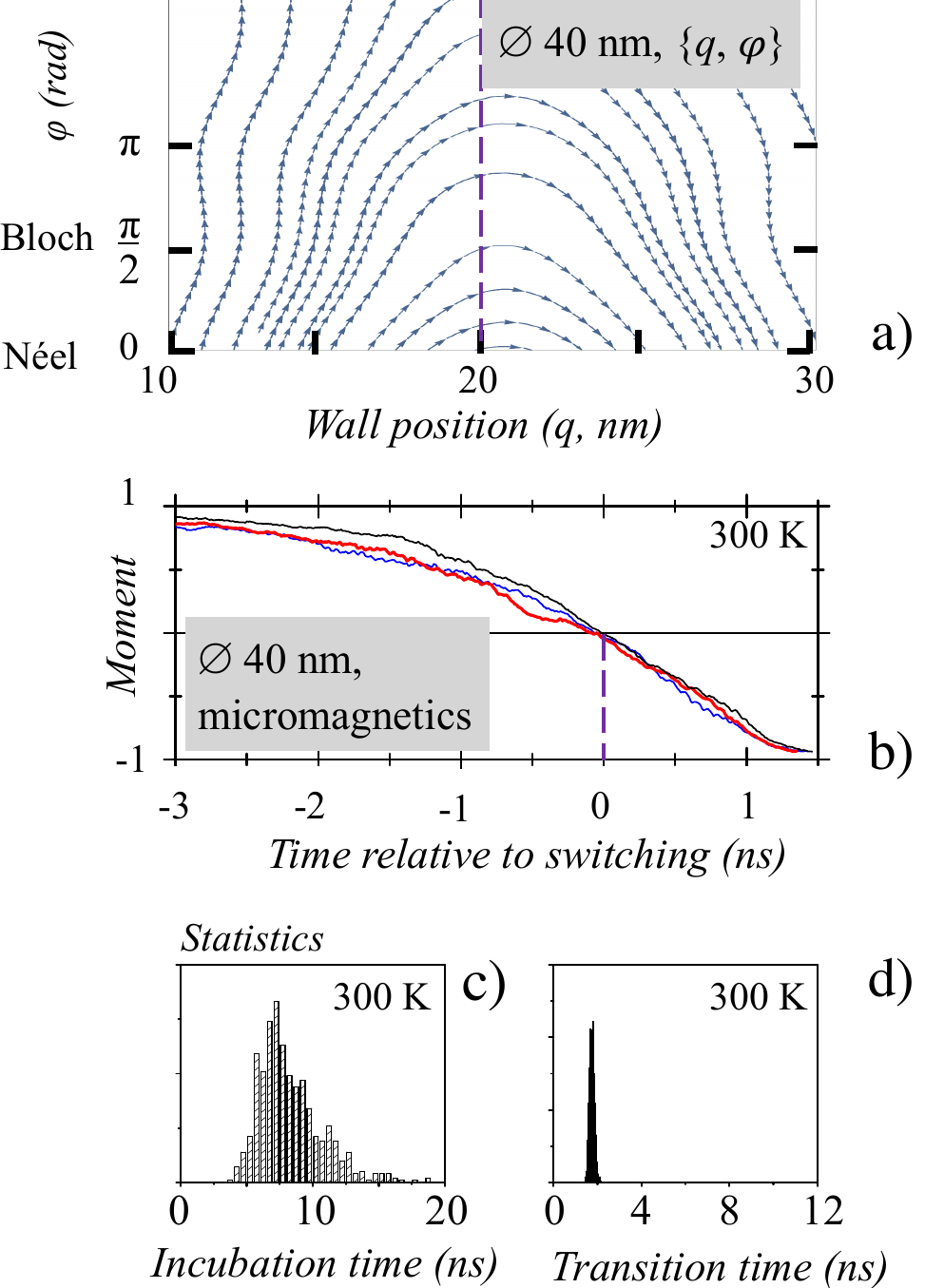}
\caption{Domain wall dynamics within a 40 nm disk at $5.4\times10^{10}~\textrm{A/m}^2$. (a) phase portrait of the wall trajectories within the $\{q,\phi\}$ collective coordinate model. (b) Three representative micromagnetic simulations of the switching at 300 K. Corresponding histograms of the incubation delay (c) and of the transition time (d).} 
\label{fig5}
\end{center}
\end{figure}


Fig.~\ref{fig4}(b) shows the phase portrait of the resulting wall trajectories in the $(q,\phi)$ space. The largely vertical flows, which takes the system from one disk edge to the other, recall that the wall motion is in the Walker regime, since many oscillations of the wall tilt $\phi$ accompany the increase in $q$. The phase portrait also shows that the number of oscillations undergone by the wall is sensitive to the initial wall tilt at small $q$, which is amplified under finite temperatures where fluctuations drive transitions between neighboring trajectories. Closer to the disk center, we observe that the Néel states ($\phi=0 [\pi]$) are associated with energy maxima, while Bloch states ($\phi=\pi /2 [\pi]$) give rise to saddle points. This representation evidences that in addition to the "switching" trajectories (i.e. wall passing from one edge to the other) there is another family of trajectories. Indeed if placing a N\'eel wall at the center, the wall is transiently held there. It needs to spiral out of this energy maximum through a lossy back-and-forth transfer of energy between communicating vessels: the position degree of freedom of the wall and the tilt degree of freedom. The region in which walls of proper tilt are transiently pinned is a "retention pond", written $\mathbb{P}$ and illustrated in Fig.~\ref{fig4}(b). 
In the conservative limit the half size of the retention pond is:
\begin{equation}
    \delta q_\mathbb{P} \approx R \sqrt{{H_{N\leftrightarrow B}}/ {H_\textrm{k,eff}}}
    \label{pondwidth}
\end{equation}

The existence of $\mathbb P$ is useful to understand the statistics of the transition time. 
The drift and the oscillation of the DW are rather independent phenomena such that two situations can occur stochastically as a wall heads to the disk center. 
If the wall avoids the vicinity of $\mathbb{P}$ [green curve in Fig.~\ref{fig4}], the $\{ q,\phi \}$ model predicts a ballistic curve. If in contrast the wall happens to tangent the retention pond (red curve), it circumvents it by performing a considerable back-and-forth motion with two pauses at either sides near the disk center [positions 1 and 4 in Fig.~\ref{fig4}(b)], in the formerly identified "central oscillation" case.  
In $\{q,\phi\}$ model, the probability of non-ballistic transition is thus correlated with the size of $\mathbb{P}$. 

If one aims at a reproducible DW propagation duration while having no handle on the wall tilt, as in the presence of thermal noise, a solution is to find a geometry \textit{without}  retention pond. In disks there exists a single "magic" disk diameter for which the pond disappears and all wall trajectories are predicted ballistic irrespective of the wall tilt. This happens when the energies of centered walls of Bloch or N\'eel characters are degenerate, i.e. when $H_{N\leftrightarrow B}=0$. With our material parameters, this diameter is 40 nm [Fig.~\ref{fig5}(a)]. One may object that this is not sufficient to warrant reproducibility of the transition time, since the $\{q,\phi\}$ model posits a uniform tilt and hence cannot account for the rare "multiple swing" trajectories that occur when the wall gets slow enough for a Bloch line to develop therein. Fortunately, the bonus of using this magic diameter is that the wall does not slow down near the disk center: this drastically reduces the probability that fluctuations can pile up and lead to the formation of a Bloch line within the wall. This optimistic conjecture was confirmed with micromagnetics [Fig.~\ref{fig5}(b-d)]. While the incubation times are still distributed for a diameter of 40 nm, the transition regimes exhibit very little variance and the distribution of transition times is particularly narrow. 

We conclude that the magic diameter ensures a repeatable wall motion independent from the tilt dynamics and immune from its fluctuations. If implemented jointly with the strategies ensuring reliable nucleation \cite{pizzini_chirality-induced_2014, lacoste_modulating_2014, bultynck_instant-spin_2018, finizio_deterministic_2019}, this strategy opens the route for reproducible switching times, which is of interest for memory applications in which write error rates could be substantially lowered. 

This work was supported by IMEC's Industrial Affiliation Program on STT-MRAM devices.

%

\end{document}